\documentclass[DIV12,11pt]{scrartcl}
\usepackage[utf8]{inputenc}
\usepackage[english]{babel}

\usepackage{amsthm,amssymb,amsmath}
\usepackage[noblocks]{authblk}

\usepackage{multirow}
\usepackage{multicol}
\usepackage{graphicx}

\usepackage{comment}
\usepackage{caption}
\usepackage{color}
\usepackage{setspace}
\newcommand{\argmax}{\operatornamewithlimits{argmax}}

\setkomafont{title}{\normalfont \LARGE \bfseries}
\setkomafont{section}{\normalfont \Large \bfseries \boldmath}
\setkomafont{subsection}{\normalfont \large \bfseries}
\setkomafont{subsubsection}{\normalfont \normalsize \bfseries}
\setkomafont{descriptionlabel}{\itshape}
\setkomafont{paragraph}{\normalfont \bfseries \boldmath}

\usepackage{fancyhdr}
\fancypagestyle{myplain}
{
  \fancyhf{}

  \fancyfoot[C]{\thepage}
}
\title{Optimization--Based Decoding Algorithms for LDPC Convolutional Codes in Communication Systems}
\vspace{-3mm}
\author[1]{Banu Kabakulak\thanks{Corresponding author.  E-mail addresses: banu.kabakulak@boun.edu.tr (B. Kabakulak), caner.taskin@boun.edu.tr (Z. C. Ta\c{s}k\i n), ali.pusane@boun.edu.tr (A. E. Pusane).}} 
\author[1]{Z. Caner Ta\c{s}k\i n}
\author[2]{Ali Emre Pusane}
\vspace{-3mm}
\affil[1]{Department of Industrial Engineering, Bo\u{g}azi\c{c}i University, \.{I}stanbul, Turkey}
\affil[2]{Department of Electrical and Electronics Engineering, Bo\u{g}azi\c{c}i University, \.{I}stanbul, Turkey}

\date{\vspace*{-2em}}

\begin{document}

\maketitle

\thispagestyle{empty}

\begin{abstract}

\vspace{-8mm}

In a digital communication system, information is sent from one place to another over a noisy communication channel. It may be possible to detect and correct  errors that occur during the transmission if one encodes the original information by adding redundant bits. Low--density parity--check (LDPC) convolutional codes, a member of the LDPC code family, encode the original information to improve error correction capability. In practice these codes are used to decode very long information sequences, where the information arrives in subsequent packets over time, such as video streams. We consider the problem of decoding the received information with minimum error from an optimization point of view and investigate integer programming--based exact and heuristic decoding algorithms for its solution. 
In particular, we consider relax--and--fix heuristics that decode information in small windows. Computational results indicate that our approaches identify near--optimal solutions significantly faster than a commercial solver in high channel error rates. Our proposed algorithms can find higher quality solutions compared with commonly used iterative decoding heuristics. 


\end{abstract}

\section{Introduction and Literature Review} \label{Introduction and Literature Review}

A digital communication system represents digital information flow from a source to sink over an unreliable environment, such as air or space. Daily communication with digital cellular phones (CDMA, GSM), high speed data modems (V. 32, V. 34), computer networks such as Internet, TV broadcasting or weather forecasting through digital satellites, image and data transmission to a space craft traveling in deep space as in the case of NASA's Pluto mission \cite{MP}, optical recording in CD-ROMs are some examples of digital communication systems. 

Since communication environments are unreliable in nature, errors may be introduced during transmission. In order to minimize the effects of these transmission errors, encoder applies certain techniques known as channel coding to add redundant bits to original information. When information reaches the receiver, decoder makes use of these redundant bits to detect and correct the errors in the received vector to obtain the original information. Work on channel coding, which started in the 1950s, has focused on turbo codes (obtained by parallel concatenation of two convolutional codes with an interleaver) and LDPC codes (described by low--density parity--check matrices). 


LDPC codes find wide application areas such as the wireless network standard (IEEE 802.11n), WiMax (IEEE 802.16e) and digital video broadcasting standard (DVB-S2) due to their high error detection and correction capabilities. LDPC code family, first proposed by Gallager in 1962, has sparse parity--check matrix representations \cite{G62}.  In the following years, LDPC codes were represented by Tanner graphs, which belong to a special type of bipartite graphs that are intensively studied in graph theory \cite{D10, BG99}.  Sparsity property of the parity--check matrix gives rise to the development of iterative message--passing decoding algorithms (such as beilef propagation, Gallager A and B algorithms) on Tanner graph with low complexity \cite{G06} -- \cite{T81}. 
Ease of the application of iterative message--passing decoding algorithms brings the advantage of low decoding latency. 

Maximum likelihood (ML) decoding is the optimal decoding algorithm in terms of minimizing error probability. Since ML decoding problem is known to be NP--hard, iterative message--passing decoding algorithms for LDPC codes are preferred in practice \cite{BMT78}. However, these heuristic decoding algorithms do not guarantee optimality of the decoded vector and they may fail to decode correctly when the graph representing an LDPC code includes cycles. 
Feldman \emph{et al.} use optimization methods and they develop linear relaxation based maximum likelihood decoding algorithms for LDPC and turbo codes in \cite{FK04, FWK05}. However, the proposed models do not allow decoding in an acceptable amount of time for codes with practical lengths.

Convolutional codes, first introduced by Elias in 1955, differ from block codes in that the encoder contains memory and the encoder outputs, at any time unit, depend both on the current inputs and on the previous input blocks \cite{E55}. Convolutional codes find application areas such as deep--space and satellite communication starting from early 1970s. They can be decoded with Viterbi algorithm, which provides maximum--likelihood decoding by dynamic programming, by dividing the received vector into smaller blocks of bits. Although Viterbi algorithm has a high decoding complexity for convolutional codes with long block lengths, it can easily implemented on hardware due to its highly repetitive nature \cite{V67, BJW10}. 
For long block lengths, sequential decoding algorithms such as Fano algorithm \cite{F63} and later stack algorithm that is developed by Zigangirov \cite{Z66} and independently by Jelinek \cite{J69} fit well. 
While Viterbi algorithm finds the best codeword, sequential decoding is suboptimal since it focuses on a certain number of likely codewords \cite{HC03}. 


LDPC Convolutional (LDPC--C) codes, introduced by J. Feltstr\"{o}m and Zigangirov in 1999, are preferred to LDPC block codes in decoding for the cases where information is obtained continuously. They can be decoded by sliding window decoders which implement iterative decoding algorithms (such as belief propagation and density evaluation) at each window \cite{JZ99}. Although LDPC--C codes provide short--delay and low--complexity in decoding, they are not in communication standards such as WiMax and DVB-S2 yet \cite{BKJ14}. 


In this study, we consider LDPC--C codes and propose optimization based sliding window decoders that can give a near optimal decoded codeword for a received vector of practical length  (approximately  $n = 4000$)  in an acceptable amount of time. The mathematical formulation and proposed decoding algorithms are explained in Section \ref{SolutionMethods}. Our proposed decoders can be used in a real--time reliable communication system since they have low decoding latency. Besides, they are applicable in settings such as deep--space communication system due to their high error correction capability. 

The rest paper is organized as follows: we define the problem in more detail in the next section. Section \ref{SolutionMethods} explains the proposed decoding techniques. We give the corresponding computational results in Section \ref{ComputationalResults}. Some concluding remarks and comments on future work appear in Section \ref{Conclusions}. 

\section{Problem Definition}\label{ProblemDefinition}

Digital communication systems transmit information from a sender to a receiver over a communication channel. Communication channels are unreliable environments, such as air, that many sender--receiver pairs share. Hence, during transmission some of the transmitted symbols can be lost or their values can change. In coding theory, information is encoded in order to overcome the occurance of such errors during the transmission. Let the information to be sent be represented by a $k$--bits long sequence $\mathbf{u} = u_{1}u_{2}... u_{k}$  $(u_{i} \in \{0, 1\})$. In order to test whether the information is sent correctly or not, parity bits are added by the encoder. This is done with a $k \times n$ generator matrix $\mathbf{G}$ through the operation $\mathbf{v} = \mathbf{uG}$ (mod 2). As a result, an $n$--bits long $(n \geq k)$ codeword  $\mathbf{v} = v_{1}v_{2}... v_{n}$  $(v_{i} \in \{0, 1\})$ is obtained. Without loss of generality, we can assume that the first  $k$ bits of the codeword are information bits, and the remaining $n-k$ bits are parity bits.  

As shown in Figure 1, when the encoded information reaches receiver as an $n$--bits long vector $\mathbf{r}$, the correctness of the vector is tested at the decoder using the parity bits. If  $\mathbf{r}$ is detected to be erroneous, the decoder attempts to determine the locations of the errors and fix them \cite{M05, BHR02}. Hence, the information $\mathbf{u}$ sent from the source is estimated as $\hat{\mathbf{u}}$ at the sink.

\begin{minipage}{\linewidth}
	\centering
	\includegraphics[width=0.65\columnwidth]{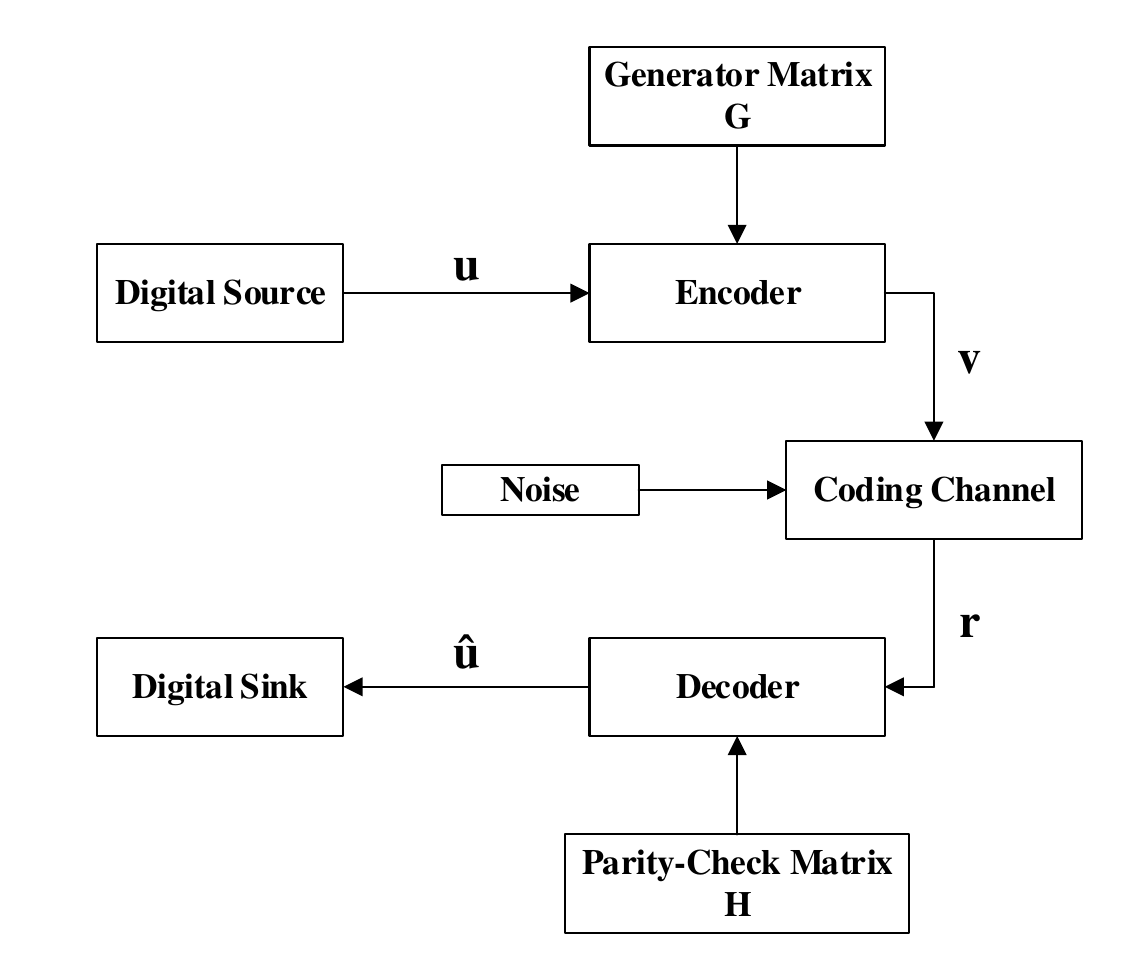}\\
	Figure 1: Digital communication system diagram 
\end{minipage}\\

There are several models used to model the noisy communication channels. In our study, we employ binary symmetric channel (BSC) model for noisy channel. As shown in Figure 2, a transmitted bit is received correctly with probability $1-p$ or an error occurs with probability $p$  \cite{M03}. 

\begin{minipage}{\linewidth}
	\centering
	\includegraphics[width=0.55\columnwidth]{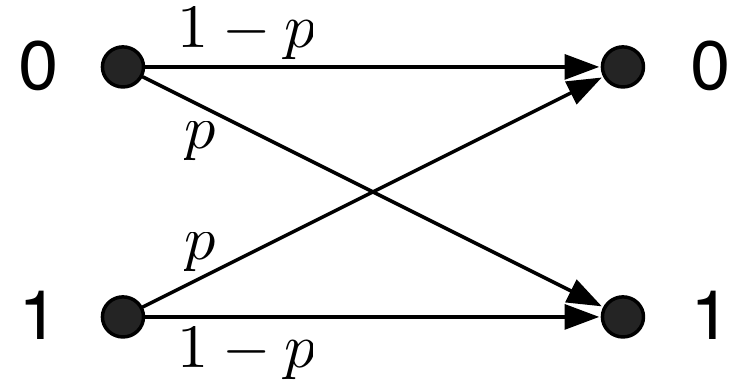}\\
	Figure 2: Binary symmetric channel 
\end{minipage}\\

In BSC, the received vector $\mathbf{r}$ includes both correct and incorrect bits. Although we do not know which bits are erroneously received, flipping the bit fixes the error when the error location is known. Hence, the aim of the decoder is to determine the error locations in BSC. 

As explained above, original information $\mathbf{u}$ is encoded with $k \times n$ generator matrix $\mathbf{G}$. Received vector $\mathbf{r}$ is decoded with a parity--check matrix $\mathbf{H}$ of dimension $(n-k) \times n$.  $(J, K)-$regular LDPC codes are member of linear block codes that can be represented by a parity--check matrix $\mathbf{H}$ with $J-$many ones at each column and  $K-$many ones at each row.  An example of a parity--check matrix from $(3, 6)-$regular LDPC code family is given in Figure 3.


\begin{minipage}{\linewidth}
	\centering
\begin{equation*}
\mathbf{H}=\begin{bmatrix}
1 & 1 &1  &1  &1  &0  &0  &0  &1 &0   \\
0 & 0 &1  &0  &0  &1  &1  &1  &1 &1   \\
1 & 0 &0  &0  &0  &1  &1  &1  &1 &1   \\
0 & 1 &0  &1  &1  &1  &1  &1  &0 &0   \\
1 & 1 &1  &1  &1  &0  &0  &0  &0 &1   \\
\end{bmatrix}
\end{equation*}
	Figure 3: A parity--check matrix from $(3, 6)-$regular LDPC code family  
\end{minipage}\\



Vectors $\mathbf{v}$ that satisfy the equation $\mathbf{v}\mathbf{H}^\textrm{T}=\mathbf{0}$ (mod 2)  are codewords.  For any original information $\mathbf{u}$,  encoded vector $\mathbf{v} = \mathbf{uG}$ (mod 2) is a codeword. The channel decoder concludes that whether the received vector $\mathbf{r}$ has changed or not by checking the value of expression $\mathbf{r}\mathbf{H}^\textrm{T}$ is equal to vector $\mathbf{0}$ in (mod 2) or not \cite{M05}.

LDPC codes can also be represented using Tanner graphs \cite{T81}. On one side of this bipartite graph, there are $n$ variable nodes standing for $n$ codeword symbols of the code and on the other side of the bipartite graph there are $(n-k)$ check nodes corresponding to $(n-k)$ parity--check equations defined by each row of the $\mathbf{H}$ matrix. Here, $\mathbf{H}$ matrix is the bi--adjacency matrix of Tanner graph. 
This representation of LDPC codes brings the advantage of applying the iterative decoding and other decoding algorithms easily. Figure 4 shows Tanner graph representation of $\mathbf{H}$ matrix defined in Figure 3. 

\begin{minipage}{\linewidth}
	\centering
	\includegraphics[width=0.65\columnwidth]{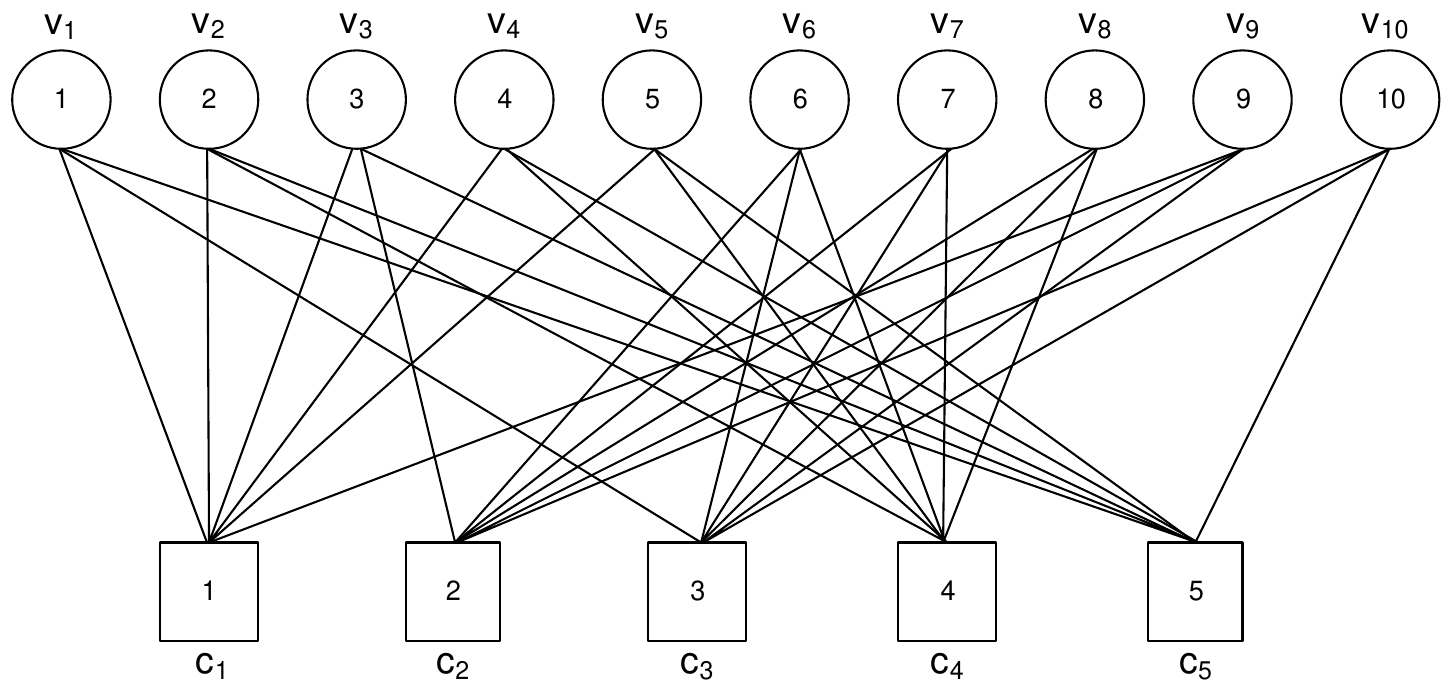}\\
	Figure 4: Tanner graph representation of the parity--check matrix given in Figure 3
\end{minipage}\\

In our study, we focus on LDPC--C codes. LDPC--C codes 
 divide the original information into smaller blocks and decode each block by considering the previous blocks \cite{JZ99}. In the code, the nonzero elements are located on the diagonal as a ribbon and the code has infinite dimension. As given in Figure 5 below, an LDPC--C code consists of  $m_\text{s}$--many  small parity--check matrices at each column, where  $m_\text{s}$ parameter represents the width of the ribbon. The diagonal pattern is obtained by shifting the columns down as the dimension increases. 

\begin{minipage}{\linewidth}
	\centering
\begin{equation*}
\mathbf{H}=\begin{bmatrix}
\mathbf{H}_{0}(1)  &       &      &   &   \\
\mathbf{H}_{1}(1)&\mathbf{H}_{0}(2)  &      &  &    \\
\vdots  &\mathbf{H}_{1}(2)  &\ddots&  &     \\
\mathbf{H}_{m_\text{s}}(1)&\vdots  &\ddots&\mathbf{H}_{0}(L)  &\\
&\mathbf{H}_{m_\text{s}}(2)&\ddots&\mathbf{H}_{1}(L)  & \ddots \\
&  &  &\vdots  & \ddots \\
& &  &\mathbf{H}_{m_\text{s}}(L) & \ddots \\
\end{bmatrix}
\end{equation*}
Figure 5: Generic structure of an LDPC--C code 
\end{minipage}\\

These codes find application areas such as satellite communication and video streams where the information is received continuously. Finite dimension LDPC--C codes, namely terminated LDPC--C codes, can be obtained by limiting the dimension of LDPC--C code by specifying a finite row or column size \cite{KRU11}. 

In Figure 6, an example of (3, 6)--regular terminated LDPC--C code obtained by limiting the row size is given.

\begin{minipage}{\linewidth}
	\centering
	\onehalfspacing
\begin{equation*}
\footnotesize
\mathbf{H}=\begin{bmatrix}
0 1 0 0 0 0 0 0 0 0 0 0 0 0 0 0 0 0 0 0 0 0 0 0  0 0 0 0 0 0 0 0 0 0 0 0 \\
1 0 1 0 0 0 0 0 0 0 0 0 0 0 0 0 0 0 0 0 0 0 0 0  0 0 0 0 0 0 0 0 0 0 0 0  \\
0 1 0 1 0 1 0 0 0 0 0 0 0 0 0 0 0 0 0 0 0 0 0 0  0 0 0 0 0 0 0 0 0 0 0 0 \\
1 0 1 0 1 0 0 1 0 0 0 0 0 0 0 0 0 0 0 0 0 0 0 0  0 0 0 0 0 0 0 0 0 0 0 0 \\
0 1 1 0 0 1 1 0 0 1 0 0 0 0 0 0 0 0 0 0 0 0 0 0  0 0 0 0 0 0 0 0 0 0 0 0 \\
1 0 0 1 1 0 0 1 1 0 0 1 0 0 0 0 0 0 0 0 0 0 0 0  0 0 0 0 0 0 0 0 0 0 0 0 \\

0 0 0 1 0 1 1 0 1 0 1 0  0 1 0 0 0 0 0 0 0 0 0 0  0 0 0 0 0 0 0 0 0 0 0 0  \\
0 0 0 0 1 0 0 1 0 1 0 1  1 0 1 0 0 0 0 0 0 0 0 0  0 0 0 0 0 0 0 0 0 0 0 0 \\
0 0 0 0 0 0 1 0 0 1 1 0  0 1 0 1 0 1 0 0 0 0 0 0  0 0 0 0 0 0 0 0 0 0 0 0 \\
0 0 0 0 0 0 0 0 1 0 0 1  1 0 1 0 1 0 0 1 0 0 0 0  0 0 0 0 0 0 0 0 0 0 0 0 \\
0 0 0 0 0 0 0 0 0 0 1 0  0 1 1 0 0 1 1 0 0 1 0 0  0 0 0 0 0 0 0 0 0 0 0 0 \\
0 0 0 0 0 0 0 0 0 0 0 0  1 0 0 1 1 0 0 1 1 0 0 1  0 0 0 0 0 0 0 0 0 0 0 0 \\

0 0 0 0 0 0 0 0 0 0 0 0 0 0 1 0 1 1 0 1 0 1 0 0  0 1 0 0 0 0 0 0 0 0 0 0 \\
0 0 0 0 0 0 0 0 0 0 0 0 0 0 0 0 1 0 0 1 1 0 0 1  1 0 1 0 0 0 0 0 0 0 0 0 \\
0 0 0 0 0 0 0 0 0 0 0 0 0 0 0 0 0 0 1 0 0 1 1 0  0 1 0 1 0 1 0 0 0 0 0 0 \\
0 0 0 0 0 0 0 0 0 0 0 0 0 0 0 0 0 0 0 0 1 0 0 1  1 0 1 0 1 0 0 1 0 0 0 0 \\
0 0 0 0 0 0 0 0 0 0 0 0 0 0 0 0 0 0 0 0 0 0 1 0  0 1 1 0 0 1 1 0 0 1 0 0 \\
0 0 0 0 0 0 0 0 0 0 0 0 0 0 0 0 0 0 0 0 0 0 0 0  1 0 0 1 1 0 0 1 1 0 0 1 \\
\end{bmatrix}
\end{equation*}
Figure 6: A (3, 6)--regular terminated LDPC--C code
\end{minipage}\\
 
(3, 6)--regular structure cannot be seen for the first and the last parts of the code. For example, the number of ones for the first five rows of the (3, 6)--regular code in Figure 6 is less than 6. Similarly, number of ones is less than 3 in the last nine columns of the code. 
One can observe the (3, 6)--regular structure for the intermediary rows and columns. 

The repeating structure of LDPC--C codes allow the application of sliding window decoding approaches which use iterative decoding algorithms (such as belief propagation, density evaluation, Gallager A and B) at each window \cite{KRU11}. Although iterative decoding algorithms are easily applicable, they cannot guarantee that the solution is near optimal. They may even fail to decode if the received vector includes errors.


Our goal in this study is to develop algorithms to decode a finite length received vector with terminated LDPC--C codes on BSC. Then we generalize these decoding algorithms to decode practically infinite length received vectors with LDPC--C codes. Our proposed decoders can give a near optimal feasible decoding for any real sized received vector in acceptable amount of time.

\section{Solution Methods}\label{SolutionMethods}

We propose three different sliding window decoders for terminated LDPC--C codes and a sliding window decoder for LDPC--C codes. The terminology used in this paper is summarized in Table \ref{tab:lop}. 

\begin{onehalfspace}
\begin{center}
\captionof{table}{List of symbols}
    \label{tab:lop}
\begin{tabular}{l l} 
\hline
\multicolumn{2}{c}{\textit{Parameters}} \\
\cline{1-2}
$k$    & length of the original information  \\
$\mathbf{G}$ & generator matrix \\
$\mathbf{H}$ & parity-check matrix \\
$\mathbf{\hat{y}}$ & received vector \\
$n$  & length of the encoded information, \# of columns in $\mathbf{H}$   \\
$p$  & error probability in BSC   \\
$m$ & \# of rows in base permutation matrix \\
$m_s$      & width of the ribbon of an LDPC--C code  \\
$C$ & set of check nodes \\
$V$ & set of variable nodes \\
$w$  & height of the window  \\
$h_s$ & horizontal step size   \\
$v_s$ & vertical step size   \\
$r$ & $h_s / v_s$ ratio   \\
\hline
\multicolumn{2}{c}{\textit{Decision Variables}} \\
\cline{1-2}
$f_i$ & $i$th bit of the decoded vector \\
$k_j$ & an auxiliary integer variable \\
\hline
\end{tabular}
\end{center}
\end{onehalfspace}

\subsection{Mathematical Formulation} \label{MathematicalFormulation}

The decoding problem of a terminated LDPC--C code can be represented with Exact Model (EM) which is  given in \cite{KD10}. The columns and rows of a $(n-k) \times n$ parity-check matrix $\mathbf{H}$ of a binary linear code can be represented with index sets $V = \{1, ..., n\}$ and $C = \{1, ..., n-k\}$, respectively.
In EM, $H_{ij}$ is the $(i, j)-$entry of parity-check matrix $\mathbf{H}$, $f_i$ is a binary variable denoting the value of the $i$th code bit and $k_j$ is an integer variable. Here, $\mathbf{\hat{y}}$ is the received vector. \\ 

\textbf{Exact Model:}
\begin{align}
& \hspace{-3pt}  \min\hspace{5pt} \sum_{i: \hat{y}_i=1}(1- f_i) + \sum_{i: \hat{y}_i=0} f_i \label{Hamming}\\
&\hspace{25pt}\mbox{s.t.} \nonumber  \\  
&\hspace{25pt}\sum_{i \in V}H_{ij}f_i = 2k_j,  \ \forall j \in C \label{sumtoeven}\\
&\hspace{25pt} f_i \in \{0, 1\}, \ \forall i \in V, \label{f_vars}\\ 
&\hspace{25pt}  k_{j} \geq 0, \ k_{j} \in  \mathbb{Z},\ \forall j \in C. \label{k_vars}
\end{align}

Constraints (\ref{sumtoeven}) guarantee that the decoded vector $\mathbf{f}$ satisfies the equality $\mathbf{f} \mathbf{H}^\textrm{T}=\mathbf{0} \ \text{(mod 2)}$. The objective (\ref{Hamming}) minimizes the Hamming distance between the decoded vector $\mathbf{f}$ and the received vector $\mathbf{\hat{y}}$. That is, the aim is to find the nearest codeword to the received vector. Constraints (\ref{f_vars}) and (\ref{k_vars}) set the binary and integrality restrictions on decision variables $\mathbf{f}$ and $\mathbf{k}$, respectively.

Since EM is an integer programming formulation, it is not practical to obtain an optimal decoding using commercial solver for real--sized (approximately $n= 4000$) terminated LDPC--C codes. This can be seen from the computational experiments in Section \ref{ComputationalResults}. Instead, we will look at terminated LDPC--C code in small windows and solve limited models at each window. 

Note that LDPC--C code decoding problem cannot be represented as a compact mathematical formulation since this would require infinite number of decision variables $f_i$ and constraints.  

\subsection{LDPC--C Code Generation} \label{SC--ConvolutionalCodeGeneration}

We implement the terminated LDPC--C code generation scheme given in \cite{AP13} which is also explained in this section. We generate a terminated LDPC--C code with the help of a base permutation matrix. 
As shown in Figure 7,  by randomly permuting the columns of an $s \times s$ identity matrix $ \mathbf{I}_s$, we can obtain a (5, 10)--regular base permutation matrix of dimension $(m, 2m)$ where $m = 5 \times s$. Regularity of the matrix is provided through augmenting identity matrices 10 times at each row and 5 times at each column. 
In Figure 7, $ \mathbf{I}_s^{i}$ represents the $i$th randomly permuted identity matrix. 

\begin{minipage}{\linewidth}
	\centering
\begin{equation*}
\mathbf{H}_{base}=\begin{bmatrix}
\mathbf{I}_s^{1}  & \mathbf{I}_s^{2}  & \mathbf{I}_s^{3}  & \mathbf{I}_s^{4}   & \mathbf{I}_s^{5}  & \mathbf{I}_s^{6}  & \mathbf{I}_s^{7}  & \mathbf{I}_s^{8}   & \mathbf{I}_s^{9} & \mathbf{I}_s ^{10} \\
\mathbf{I}_s^{11}  & \mathbf{I}_s^{12}  & \mathbf{I}_s^{13}  & \mathbf{I}_s^{14}   & \mathbf{I}_s^{15}  & \mathbf{I}_s^{16}  & \mathbf{I}_s^{17}  & \mathbf{I}_s^{18}   & \mathbf{I}_s^{19} & \mathbf{I}_s ^{20} \\
\mathbf{I}_s^{21}  & \mathbf{I}_s^{22}  & \mathbf{I}_s^{23}  & \mathbf{I}_s^{24}   & \mathbf{I}_s^{25}  & \mathbf{I}_s^{26}  & \mathbf{I}_s^{27}  & \mathbf{I}_s^{28}   & \mathbf{I}_s^{29} & \mathbf{I}_s ^{30} \\
\mathbf{I}_s^{31}  & \mathbf{I}_s^{32}  & \mathbf{I}_s^{33}  & \mathbf{I}_s^{34}   & \mathbf{I}_s^{35}  & \mathbf{I}_s^{36}  & \mathbf{I}_s^{37}  & \mathbf{I}_s^{38}   & \mathbf{I}_s^{39} & \mathbf{I}_s ^{40} \\
\mathbf{I}_s^{41}  & \mathbf{I}_s^{42}  & \mathbf{I}_s^{43}  & \mathbf{I}_s^{44}   & \mathbf{I}_s^{45}  & \mathbf{I}_s^{46}  & \mathbf{I}_s^{47}  & \mathbf{I}_s^{48}   & \mathbf{I}_s^{49} & \mathbf{I}_s ^{50} \\
\end{bmatrix}
\end{equation*}
	Figure 7: (5, 10)--regular base permutation matrix
\end{minipage}
\vspace{3mm}

Then, we split the base permutation matrix into two matrices, namely lower triangular $\mathbf{A}$ and upper triangular $\mathbf{B}$ as shown in Figure 8. 

\begin{minipage}{\linewidth}
	\centering
	\includegraphics[width=0.5\columnwidth]{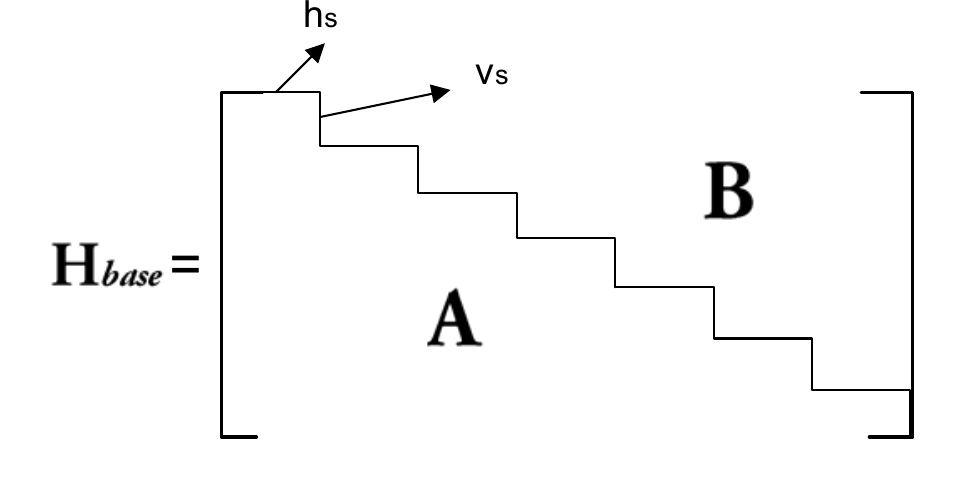}\\
	Figure 8:  $\mathbf{A}$ and $\mathbf{B}$ matrices
\end{minipage}\\

\vspace{4mm}

We divide $\mathbf{H}_{base}$ with a horizontal step length $h_s$ and a vertical step length $v_s$. One can observe that when $\mathbf{H}_{base}$ is (5, 10)--regular, its dimension is $(m, 2m)$ for some $m$. Then, $r = h_s / v_s  = 2$, since there is a number $c$ such that $h_s c = 2m$ and $v_s c  = m$.

\begin{minipage}{\linewidth}
	\centering
	\includegraphics[width=0.82\columnwidth]{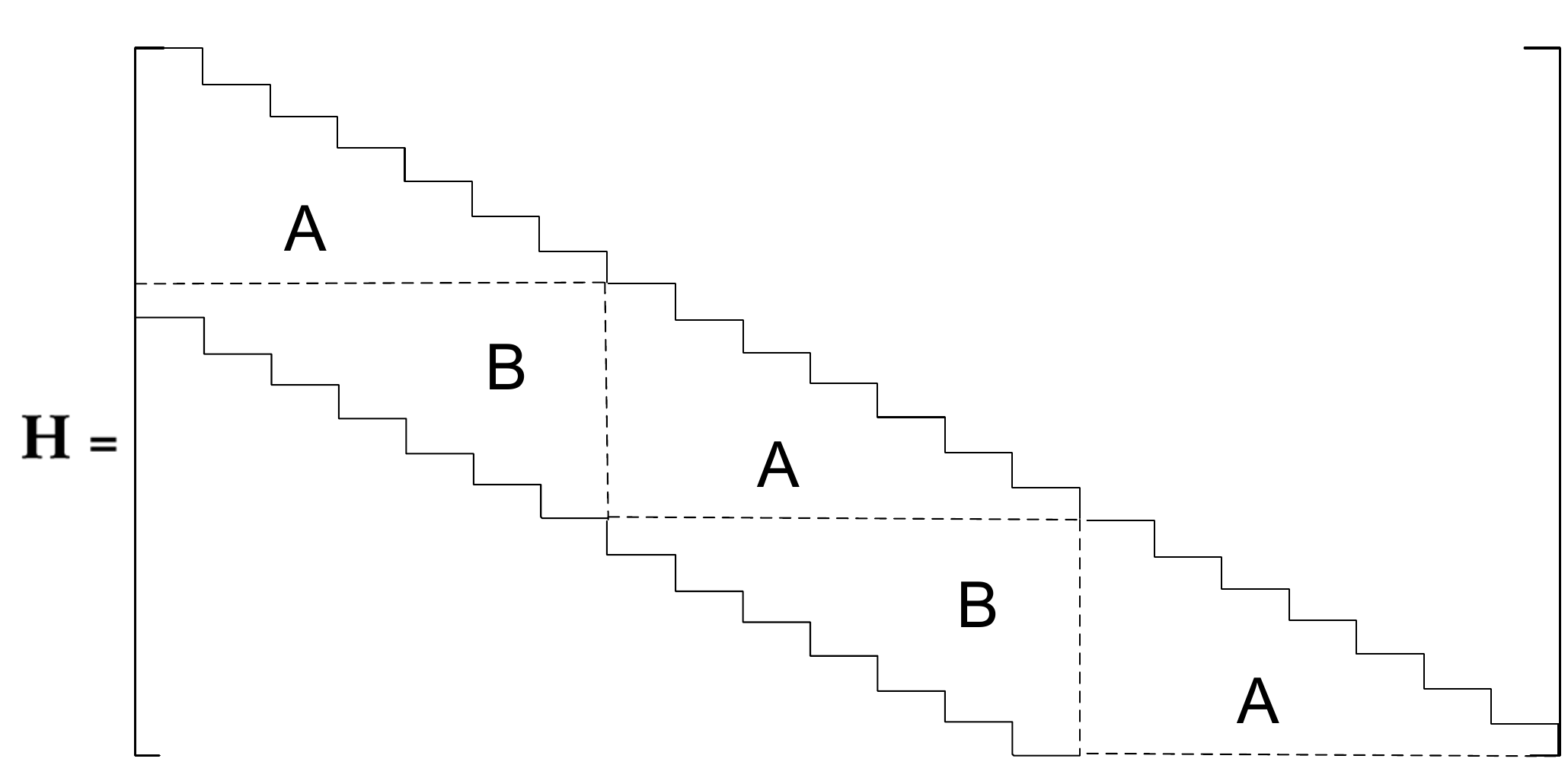}\\
	Figure 9:  (5, 10)--regular terminated LDPC--C code
\end{minipage}\\

\vspace{4mm}

Then, these  $\mathbf{A}$ and $\mathbf{B}$ matrices are repeatedly located until the desired terminated LDPC--C code size is obtained. After $t$--many repetitions, terminated LDPC--C code has size  $(t m, 2t m)$ as shown in Figure 9. The ribbon size is $m_s = m + v_s$ for such a code.

\subsection{Sliding Window Decoders} \label{SlidingWindowDecoders}

Sliding window decoders in practical applications make use of special structure of the LDPC--C codes \cite{LSC+10, CPB+06}. As explained in Section \ref{ProblemDefinition}, LDPC--C codes have all nonzero entries on a ribbon, with width $m_s$, that lies on the diagonal. Then, one can consider a window on the LDPC--C code with height $w$ 
and decode the received vector partially. Decoding of the received vector proceeds iteratively by sliding the window $h_s$ units horizontally and $v_s$ units vertically. 


In sliding window decoders, we can pick window row size $w  > m_s$ and column size larger or equal to $r w$  where $r = h_s / v_s$. For the rows of the LDPC--C code corresponding to the window, all entries in the columns after the window are zero with this window dimension selection.

Algorithm 1 explains the main steps of a generic sliding window decoder. Part of the received vector corresponding to the current window is decoded with an algorithm. Hence, performance of a sliding window decoder depends on how fast and correctly the windows are decoded. As we mention in Section \ref{ProblemDefinition}, if we implement Gallager A or B algorithm for windows, decoded vector may not close to the original information. We investigate the performance of Gallager A and B algorithms in sliding window decoder with computational experiments in Section \ref{ComputationalResults}. 

\vspace{2mm}
\begin{center}
$ 
\onehalfspacing
\begin{tabular}{p{11cm}}
\textbf{Algorithm 1:} (Generic Sliding Window) \\
\hline
\textbf{Input:} Received vector $\mathbf{\hat{y}}$, Binary code $\mathbf{H}$  \\
\hline
\vspace{-4mm}
\begin{enumerate}
\item Decode the current window with an algorithm
\item Move the window $h_s$ units horizontally, $v_s$ units vertically
\item Fix the decoded values of $h_s$--many leaving bits
\item \textbf{If} all bits decoded,  \textbf{Then}  STOP, \textbf{Else} go to Step 1
\end{enumerate}\\
\hline
\textbf{Output:} A decoded vector/codeword \\
\hline
\end{tabular}
$\end{center}

\vspace{4mm}




In our approach, we solve each window  with EM formulation that is written for the decision variables and constraints within the window. At each iteration, $h_s$--many bits and $v_s$--many constraints leave the window. Exiting bits are decoded in the previous window and can be fixed to their decoded values in the proceeding iterations. The decoded bits will affect the upcoming bits by appearing as a constant in the constraints (\ref{sumtoeven}). Our sliding window decoding algorithm has main steps that are given in Algorithm 2.

\vspace{2mm}
\begin{center}
$ 
\onehalfspacing
\begin{tabular}{p{11cm}}
\textbf{Algorithm 2:} (Sliding Window) \\
\hline
\textbf{Input:} Received vector $\mathbf{\hat{y}}$, Binary code $\mathbf{H}$  \\
\hline
\vspace{-4mm}
\begin{enumerate}
\item Solve EM for the current window 
\item Move the window $h_s$ units horizontally, $v_s$ units vertically
\item Fix the decoded values of $h_s$--many leaving bits
\item Update constraints  (\ref{sumtoeven}) with the fixed bits
\item \textbf{If} all bits decoded,  \textbf{Then}  STOP, \textbf{Else} go to Step 1
\end{enumerate}\\
\hline
\textbf{Output:} A decoded codeword\\
\hline
\end{tabular}
$\end{center}

\vspace{4mm}

It is possible to apply different strategies in window dimension selection and window solution generation. This gives rise to our four different sliding window decoders, i.e. complete window, finite window and repeating windows decoders for terminated LDPC--C codes and an LDPC--C code decoder, that are explained in the next sections.

\subsubsection{Complete Window (CW) Decoder} \label{CWDecoder}

Complete window (CW) decoder requires that binary code has finite dimension. Hence, it is applicable only for terminated LDPC--C codes. In CW, the window height is $w$ and width is $n$ (the length of the received vector $\mathbf{\hat{y}}$). This means in a window we have $w$--many constraints and $n$--many bits as $f_i$ decision variables.   

We consider two diffrent ways in window decoding. In the first approach, i.e. Some Binary CW (SBCW), we restrict the first undecoded $h_s$ bits of the window to be binary and relax the bits coming after those as continuous variables. As an example, when we solve the first window of the code in Figure 10, first $h_s$ bits (corresponding to the dotted rectangle) are binary and we relax all the remaining bits as continuous. When we move to the next window by shifting the window $v_s$ units down, first $h_s$ bits have been fixed to their decoded values, the next $h_s$ bits are set to be binary and the bits coming after are continuous variables. The decoder proceeds in this fashion.

\vspace{3mm}

\begin{minipage}{\linewidth}
	\centering
	\includegraphics[width=1.0\columnwidth]{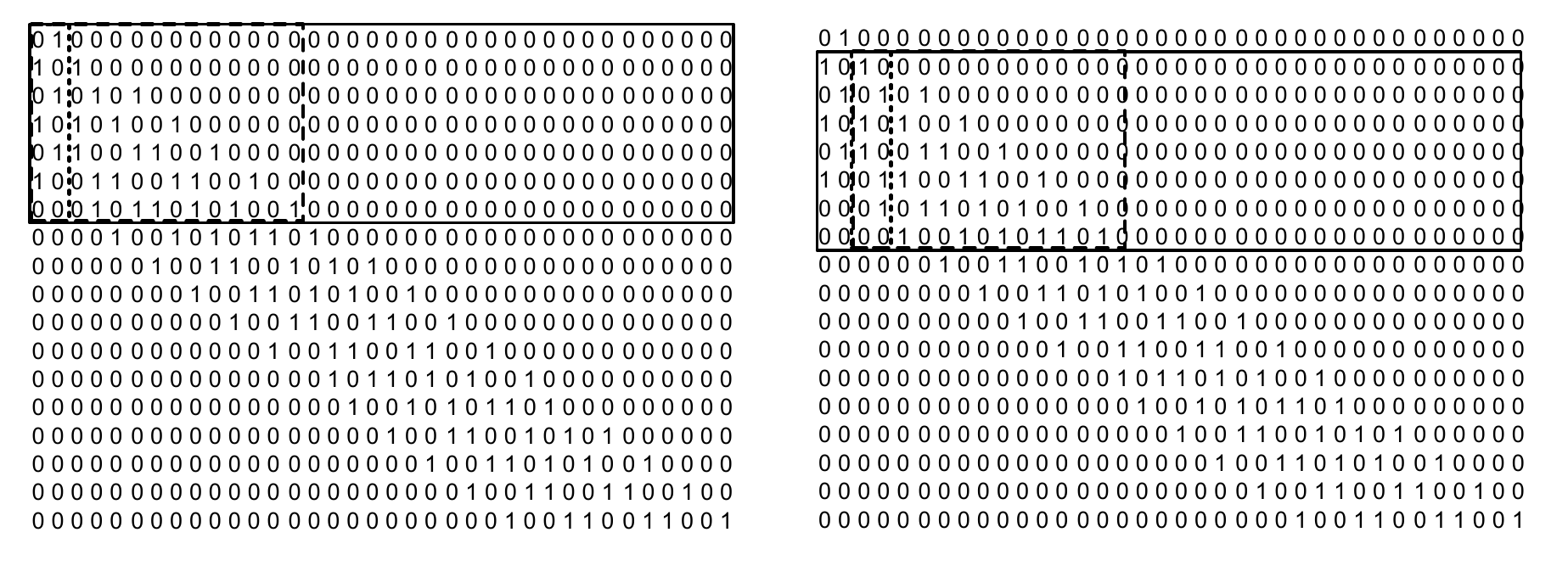}\\
	Figure 10: Sliding window in CW decoder
\end{minipage}\\

\vspace{3mm}

One can see that the dashed rectangle in Figure 10 covers all nonzero entries in the window. From this observation as a second approach, i.e. All Binary CW (ABCW), we consider to force the first undecoded $(r w)$--many bits (corresponding to the dashed rectangle) of the window to be binary and the ones after these are continuous. As we move to the next window, $h_s$--many bits are fixed and the dashed rectangle shift to right $h_s$ units. Moving from one window to the other requires removing first $v_s$--many constraints and including new $v_s$--many constraints.

The method of fixing some of the decision variables and relaxing some others is known as Relax--and--Fix heuristic in the literature \cite{W98, FMM14}. In general, fixing the values of the variables may lead to infeasibility in the next iterations. However, we do not observe such a situation in our computational experiments 
when we pick the window that is sufficiently large to cover all nonzero entries for the undecoded bits in the corresponding rows. 
We can observe that a window of size $w \times (r w)$ (dashed rectangle) can cover the undecoded nonzero entries.





\subsubsection{Finite Window (FW) Decoder}  \label{FWDecoder}

In finite window (FW) decoder, we have smaller window of size $w \times (r w)$. That is we have $w$--many constraints and $ (r w)$--many  $f_i$ decision variables. At each iteration, after solving EM model for the window, we fix first $h_s$--many bits and slide the window. In Some Binary FW (SBFW) decoder, we restrict first $h_s$--many bits to be binary and relax the rest as continuous. For All Binary FW (ABFW) method, all  $(r w)$--many bits are binary variables. 

\begin{minipage}{\linewidth}	
	\centering
	\includegraphics[width=1.0\columnwidth]{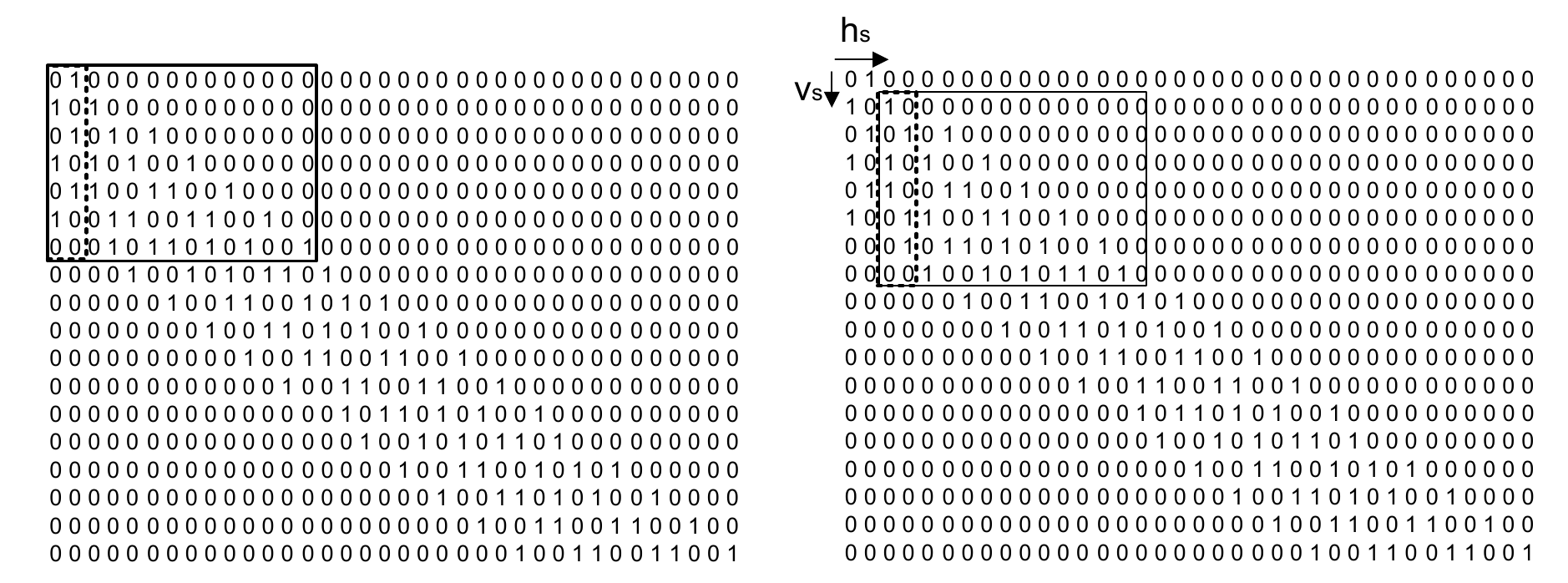}\\
	Figure 11: Sliding window in FW decoder
\end{minipage}\\
\vspace{3mm}

The window position can be seen in Figure 11 as the window slides. The previous decoded bits appear as a constant in constraints (\ref{sumtoeven}) of EM formulation for the current window. In FW, we store only one window model. This means we are storing $w$--many constraints and $(r w)$--many $f_i$ decision variables in the memory at a time. 

As we move from one window to the other, we remove $h_s$--many decision variables and introduce $h_s$--many new ones. Also, we remove $v_s$--many constraints and add $v_s$--many new constraints. 

\subsubsection{Repeating Windows (RW) Decoder}  \label{RWDecoder}

As explained in Section \ref{SC--ConvolutionalCodeGeneration}, a terminated LDPC--C code is obtained by repetitively locating $\mathbf{A}$ and $\mathbf{B}$ matrices. As can be seen in Figure 12, a window will come out again after $m$--iterations, where $m$ is the number of rows in $\mathbf{H}_{base}$. 


\begin{minipage}{\linewidth}	
	\centering
	\includegraphics[width=1.0\columnwidth]{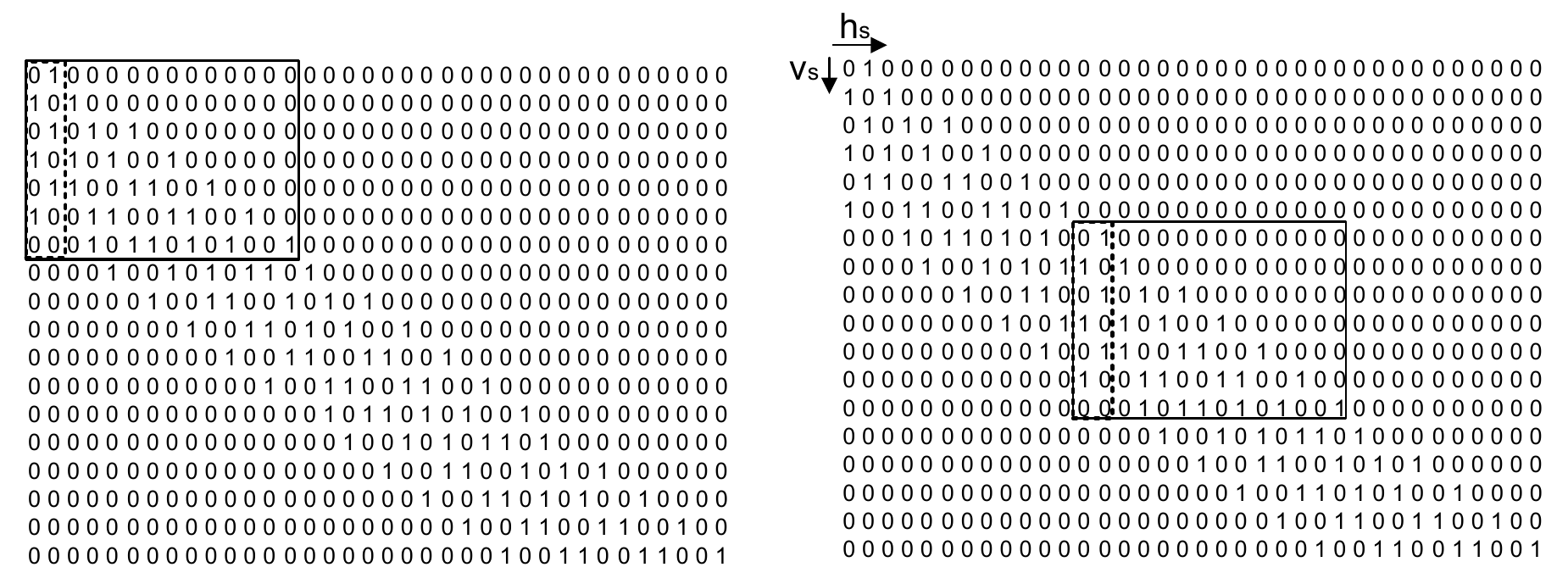}\\
	Figure 12: Sliding window in RW decoder
\end{minipage}\\

\vspace{3mm}

This means that there are $m$--many different windows. However, the first and the $(m + 1)$st windows still differ from each other in terms of their EM formulation. That is the constant term in constraints (\ref{sumtoeven}) and the objective function coefficients change but the coefficients of the decision variables stay the same. Hence, we store $m$--many window models and when its turn comes we solve the window after updating the constant term and the objective function. 

Assuming that a window is of size $w \times (r w)$, having $m$--many window models requires to store $(m w)$--many constraints and $(m r w)$--many $f_i$ decision variables in the memory. However, we do not need to add or remove constraints and decision variables. FW decoder has the burden of add/remove operations and the advantage of low memory usage. On the other hand, RW decoder directly calls the window models on the expense of memory.


In Some Binary RW (SBRW) only first $h_s$--many bits are binary, whereas All Binary RW (ABRW) has all  $(r w)$--many bits as binary variables.

\subsubsection{LDPC Convolutional Code (CC) Decoder}  \label{CCDecoder}

The decoders CW, FW and RW assume that we are given a finite dimensional code that can be represented by a $\mathbf{H}$ matrix. Hence, they are applicable for terminated LDPC--C codes. However, as explained in Section \ref{ProblemDefinition}, LDPC--C codes are practically infinite dimensional codes and cannot be represented by a compact $\mathbf{H}$ matrix on computer. On the other hand, they are generated from $\mathbf{A}$ and $\mathbf{B}$ matrices. Therefore, we can store a part of LDPC--C code as given in Figure 13 that includes the required information.  

\begin{minipage}{\linewidth}	
	\centering
	\includegraphics[width=0.4\columnwidth]{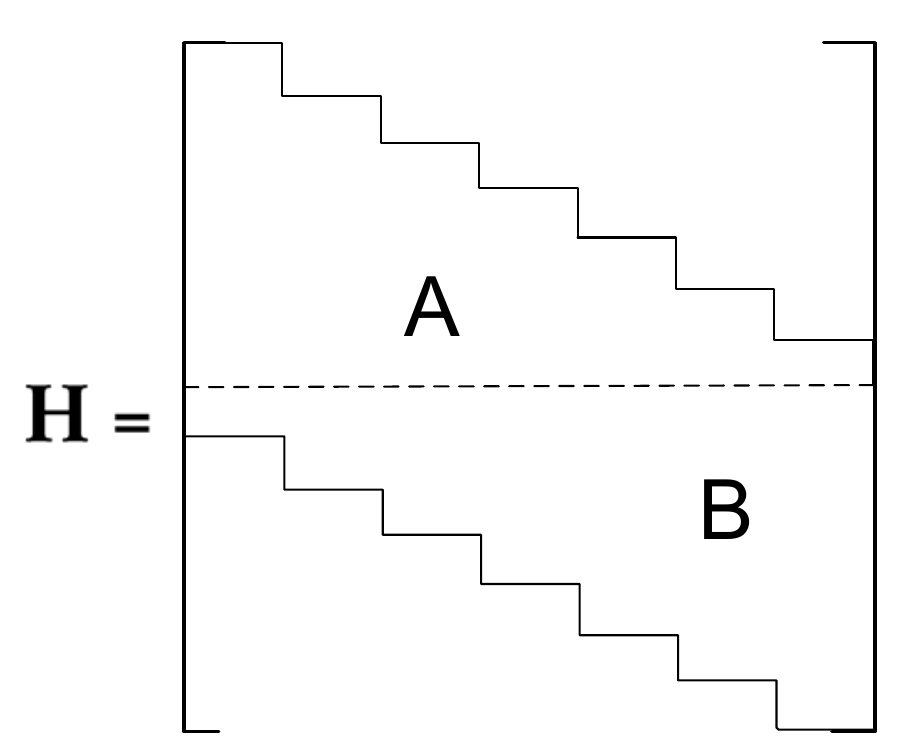}\\
	Figure 13: A part of LDPC--C code
\end{minipage}\\

\vspace{3mm}

With this part of the LDPC--C code, we can represent the $(i, j)$th entry of the code with a function. Hence, we can represent the current window model using this small matrix. This allows the application of FW and RW decoders to LDPC--C codes. Note that our CW decoder is not applicable to CC, since it takes into account all bits of the received vector. 


\section{Computational Results} \label{ComputationalResults}


The computations have been carried out on a computer with 2.6 GHz Intel Core  i5-3230M processor and 4 GB of RAM working under Windows 10 Professional.

In our computational experiments, we evaluate the performance of our sliding window decoders. In our decoders, the number of the constraints and decision variables in EM formulation limited with the size of the window. We make use of CPLEX 12.6.0 to solve EM for the current window (see Step 1 of Algorithm 2). We compare the performance of our sliding window decoders with Exact Model Decoder (EMD). In EMD, EM formulation includes all constraints and decision variables corresponding to terminated LDPC--C code. That is, for a terminated LDPC--C code of size $(n/2, n)$ we have $n/2-$many constraints (\ref{sumtoeven}) and $n-$many $f_i$ decision variables in EM. We again utilize CPLEX for solving EM of EMD. 

\newpage

\begin{onehalfspace}
\begin{center}
\footnotesize
\captionof{table}{Summary of methods}
    \label{tab:som}
\begin{tabular}{ccccccccc} 
\hline
Method & \# CPLEX   & Move to  & Window  & \multicolumn{2}{c} {\# Vars}  & & \multicolumn{2}{c} {\# Vars}  \\
\cline{5-6}  \cline{8-9} 
  & Models  &  Next Window  & Size & $k_j$ & $f_i$  & & & int./binary/cont.  \\
\hline
CW  & 1 & Delete/add $v_s$--const. & $w\times n$ & $w$ & $n$ & &  SB & $w / h_s / (n- h_s)$ \\
 &   & Update const. (\ref{sumtoeven}) &  &  & & &AB & $w / rw / (n-rw)$  \\
FW  & 1 & Delete/add $v_s$--const., $h_s$--vars & $w\times rw$ & $w$ & $rw$ & &  SB & $w / h_s / (rw- h_s)$ \\
 &   & Update const. (\ref{sumtoeven}) &  &  & & &AB & $w / rw / 0$  \\
RW  & $m$ & Update obj. func. coeffs & $w\times rw$ & $mw$ & $mrw$ & &  SB & $mw / mh_s / m(rw- h_s)$ \\
 &   & Update const. (\ref{sumtoeven}) &  &  & & &AB & $mw / mrw / 0$  \\
EMD  & 1 & --- & $(n/2)\times n$ & $(n/2)$ & $n$ & &  --- & $(n/2) / n / 0$ \\
\hline
\end{tabular}
\end{center}
\end{onehalfspace}

We summarize the solution methods in Table \ref{tab:som}. ``\# CPLEX Models" gives the number of CPLEX models stored in the memory. One needs to carry out operations given in ``Move to Next Window" column when sliding the window of size ``Window Size". In ``\# Vars" columns, we list the number of $k_j$ and $f_i$ decision variables and also give the number of integer, binary and continuous decision variables stored in the memory for SB and AB approaches of the methods. CC decoder is not listed in Table \ref{tab:som}, since it is the application of FW and RW decoders to practically infinite dimensional codes. 


\begin{onehalfspace}
\begin{center}
\captionof{table}{List of computational parameters}
    \label{tab:locp}
\begin{tabular}{l l} 
\hline
\multicolumn{2}{c}{\textit{Parameters}} \\
\cline{1-2}
$n$  & 1200, 3600, 6000, 8400, 12000   \\
$p$  & 0.02 (low), 0.05 (high) \\
$m$ & 150 \\
$w$  &  $m$ + 1 (small), $\frac{3 m }{2}+ 1$ (large) \\
$h_s$ & 2  \\
$v_s$ & 1   \\
\hline
\end{tabular}
\end{center}
\end{onehalfspace}

A summary of the parameters that are used in the computational experiments are given in Table \ref{tab:locp}. We generate a base permutation matrix of size $(m , 2m) = (150, 300)$. We obtain a (5, 10)--regular terminated LDPC--C code $\mathbf{H}$ of desired dimensions from this base permutation matrix. In our experiments, we consider four different code length, i.e. $n = 1200, 3600, 6000, 8400$ for terminated LDPC--C codes. In order to test the algorithms for LDPC--C codes,  we consider a larger code length $n = 12 000$. For each code length $n$, we experiment 10 random instances and report the average values. We investigate two levels of error rate, i.e. low error $p = 0.02$ and high error $p = 0.05$. There are two alternatives for the window sizes, namely small window $w = m + 1$ and large window $w = \frac{3 m }{2}+ 1$.

In our sliding window algorithms, we solve the window models with CPLEX within 1 minute time limit. On the other hand, we set a time limit of 4000 seconds  to EMD for solving a terminated LDPC--C code instance. Since we are testing a larger code length, i.e. $n = 12 000$, for LDPC--C codes, we set a time limit of 5000 seconds to EMD to find a solution. 

\begin{onehalfspace}
\begin{center}
\captionof{table}{Performance of EMD with $p = 0.02$ and $0.05$}
    \label{tab:CPLEX--LowHighErrorSC}
\begin{tabular}{ccccccccccccc}
    \hline
 $p$ & \multicolumn{4}{c}{ $0.02$} & & \multicolumn{4}{c}{ $0.05$}  \\ \cline{2-5} \cline{7-10}
$n$  & $z$ & CPU & Gap (\%) & \# OPT & & $z$ & CPU & Gap (\%) & \# OPT\\
    \hline 
1200 & 23.9 & 0.16 & 0 & 10 & & 56.3 & 221.42 & 0 & 10 \\
3600 & 72.7 & 0.23 & 0 & 10 & & 736.0 & 1797.72 & 35.44 & 6 \\
6000 & 121.0 & 0.32 & 0 & 10 & & 1358.3  & 3890.06 & 43.37 & 4 \\
8400 & 169.9 & 0.54 & 0 & 10 & & 3623.7 & 4049.98 & 80.45 & 1 \\
\hline
12000 & 238.6 & 0.85 & 0 & 10 & & 4300.6 & 4457.34 & 70.84 & 2 \\
\hline
\end{tabular}
\end{center}
\end{onehalfspace}
\vspace{2mm}

Table \ref{tab:CPLEX--LowHighErrorSC} gives the performance of EMD under low and high error rates. The column ``$z$" shows the objection function value of the best known solution found within the time limitation. ``CPU" is the computational time in terms of seconds. ``Gap (\%)" is the relative difference between the best lower and upper bounds. ``\# OPT" is the number of instances that are solved to optimality among 10 trials. The first four rows in Table \ref{tab:CPLEX--LowHighErrorSC}  are average results for terminated LDPC--C codes. 
The last row is the average result for LDPC--C code.  As the error rate increases, EMD has difficulty in finding optimal solutions. A similar pattern is observed when the length of the received vector $n$ increases. That is, the optimality gap increases when the code gets longer as expected. 

\subsection{Terminated LDPC--C Code Results}

In this section, we discuss the results of the computational experiments of $n = 1200, 3600, 6000, 8400$ for error probabilities 0.02 and 0.05 and two levels of window size, i.e., small and large.

\newpage
\begin{onehalfspace}
\begin{center}
\footnotesize
\captionof{table}{Performances of SBCW and ABCW}
    \label{tab:SBABCW}
\begin{tabular}{ccccccccccccc}
    \hline
& &  $w$  & \multicolumn{4}{c}{small} &  & \multicolumn{4}{c}{large}  \\ \cline{4-7} \cline{9-12}
& $p$ & $n$ & $z$ & CPU & Gap (\%) & \# SOLVED &  & $z$ & CPU & Gap (\%) & \# SOLVED \\
    \hline
SB & 0.02 & 1200 &  23.9  & 5.48 &  0 & 10 &  & 23.9  & 6.31 &  0 & 10 \\
&& 3600 &  72.7  & 31.89 & 0 & 10  &  &  72.7  & 39.05 &  0 & 10 \\
&& 6000 & 121.0  & 83.77 & 0 & 10 &  &  121.0  & 97.54 &  0 & 10 \\
&& 8400 & 169.9  & 168.97 & 0 & 10 &  &  169.9  & 187.91 & 0 & 10 \\
&0.05 & 1200 &  56.3  & 26.70 & 0 & 10 &  & 107.3 & 334.28 & 9.03 & 10 \\ 
&& 3600 & 181.8  & 224.30 & 2.83 & 10 &  & 1052.1 & 1056.68 & 53.46 & 10 \\
&& 6000 & 564.2  & 1165.43 & 14.92 & 10 &  & 2504.2 & 1419.37 & 80.25 & 10  \\
&& 8400 & 2243.75  & 2188.73 & 48.27 & 10 &  & 4016.5 & 1346.33 & 89.48 & 10 \\
AB & 0.02 & 1200 & 23.9 & 6.33 & 0 & 10  &  &  23.9 & 7.65 & 0 & 10 \\
&& 3600 & 72.7 & 34.05 & 0 & 10 &  & 72.7 & 43.86 & 0 & 10 \\
&& 6000 & 121.0 & 88.51  & 0 & 10 &  & 121.0 & 107.59 & 0 & 10 \\
&& 8400 &  169.9 & 177.23 & 0 & 10 &  & 169.9 & 201.58 & 0 & 10 \\
&0.05 & 1200 & 58.3 & 17.85 & 2.59 & 10 &  & 56.3 & 57.01 & 0 & 10 \\ 
&& 3600 & 181.8 & 67.58 & 2.83 & 10 &  & 614.9 & 494.62 & 26.82 & 10 \\
&& 6000 & 392.9 & 537.07 & 16.81 & 10 &  & 1279.3 & 941.14 & 36.05 & 10 \\
&& 8400 &  533.5 & 642.54 & 17.50 & 10 &  & 3119.4 & 761.03 & 67.13 & 10 \\
\hline
\end{tabular}
\end{center}
\end{onehalfspace}
\vspace{2mm}

Table \ref{tab:SBABCW} summarizes the results for CW decoder explained in Section \ref{CWDecoder}. ``Gap (\%)" column represents the percent difference from the best known lower bound found by CPLEX while obtaining the results in Table \ref{tab:CPLEX--LowHighErrorSC}. ``\# SOLVED" column shows the number of instances that can be decoded by the method.

When $p = 0.02$, CW decoder can find optimal solutions as EMD in Table \ref{tab:CPLEX--LowHighErrorSC}. However, CW completes decoding in longer time for both SB and AB variants and both window sizes. This is since solving EM model with CPLEX (in EMD) under low error probability is easy and decoding in small windows takes longer time in CW.
When the error probability increases to $0.05$ and window size is small, we can see that CW finds better feasible solutions in shorter time than EMD (in Table \ref{tab:CPLEX--LowHighErrorSC}) for SB and AB variants. As the window size gets larger, only AB alternative gives better gap and time values compared with EMD.

In general, with high error probability AB takes shorter time and obtains better gaps than SB (see results for $p = 0.05$ in Table \ref{tab:SBABCW}, Table \ref{tab:SBABFW} and Table \ref{tab:SBABRW}). Note that this is somewhat counter intuitive since the number of binary variables in AB variant is larger than SB. However, note that AB has the advantage of being able to use the integral solution of the previous window as a starting solution of the new window. Hence, AB has more time to find a better solution in the current window within the time limit compared with SB. 


When $p = 0.05$, the performance of CW deteriorates as the window size gets larger. Solving a larger model in a window decreases the quality of the solution obtained within the time limit. Size of the window model also depends on the length of the received vector $n$. Hence, the gap values increase as $n$ increases. 

\begin{onehalfspace}
\begin{center}
\footnotesize
\captionof{table}{Performances of SBFW and ABFW}
    \label{tab:SBABFW}
\begin{tabular}{ccccccccccccc}
    \hline
& &  $w$  & \multicolumn{4}{c}{small} &  & \multicolumn{4}{c}{large}  \\ \cline{4-7} \cline{9-12}
&$p$ & $n$ & $z$ & CPU & Gap (\%) & \# SOLVED &  & $z$ & CPU & Gap (\%) & \# SOLVED \\
    \hline
SB &0.02 & 1200 & 23.9  & 10.91 & 0 & 10 &  &  23.9  & 11.90 & 0 & 10 \\
&& 3600 & 72.7  & 24.25 &  0 & 10 &  & 72.7  & 47.47 &  0 & 10 \\
&& 6000 & 121.0  & 41.98 &  0 & 10 &  & 121.0  & 74.04 & 0 & 10 \\
&& 8400 & 169.9  & 62.48 &  0 & 10 &  & 169.9 & 129.44 & 0 & 10 \\
&0.05 & 1200 &  56.3  & 15.07 & 0 & 10 &  &  56.3  & 364.67 & 0 & 10  \\ 
&& 3600 & 196.6  & 348.29 & 5.89 & 10 &  & 177.0  & 3581.46 & 0.78 & 10 \\
&& 6000 & 353.9  & 973.76 & 12.89 & 10 &  & 300.3  & 6889.76 & 0.86 & 10 \\
&& 8400 & 629.8  & 3561.43 & 26.65 & 10 &  & 427.0  & 11445.70 & 1.03 & 10 \\
AB &0.02 & 1200 & 23.9 & 6.59 & 0 & 10 &  & 23.9 & 15.80 & 0 & 10 \\
&& 3600 & 72.7 & 22.95 & 0  & 10 &  & 72.7 & 65.66 & 0 & 10 \\
&& 6000 & 121.0 & 39.22 & 0 & 10 &  & 121.0 & 114.59 & 0 & 10 \\
&& 8400 & 169.9 & 56.45 & 0 & 10 &  & 169.9 & 165.71 & 0 & 10 \\
&0.05 & 1200 & 58.3 & 21.48 & 2.59 & 10 &  & 56.3 & 72.78 & 0 & 10 \\ 
&& 3600 & 214.6 & 387.94 & 10.97 & 10 &  & 177.0 & 999.93 & 0.78 & 10 \\
&& 6000 & 368.1 & 653.55 & 16.05 & 10 &  & 300.3 & 2087.41 & 0.86 & 10 \\
&& 8400 & 617.8 & 1792.25 & 25.96 & 10 &  & 427.0 & 3061.82 & 1.03 & 10 \\
\hline
\end{tabular}
\end{center}
\end{onehalfspace}
\vspace{2mm}

Results given in Table \ref{tab:SBABFW} shows that FW (see Section \ref{FWDecoder}) can find optimal solution in all cases when $p = 0.02$. With this error probability, FW needs more time to find the optimal solution for SB and AB alternatives when the window size gets larger. The computational times are larger than EMD for both alternatives.

However, as error probability gets higher, FW can find better solutions than EMD in shorter time for SB and AB methods. AB method is faster than SB, since it can make use of integral solution found in the previous window. Note that a similar pattern also appears in CW as discussed before.
FW takes more time than CW for both SB and AB alternatives, since it needs to add and remove variables while moving to the next window position. On the other hand, the size of the window model is independent from code length $n$, hence we can find better solutions within the time limit. As a result, the gap values are better than CW decoder. 

\begin{onehalfspace}
\begin{center}
\footnotesize
\captionof{table}{Performances of SBRW and ABRW}
    \label{tab:SBABRW}
\begin{tabular}{ccccccccccccc}
    \hline
& &  $w$  & \multicolumn{4}{c}{small} &  & \multicolumn{4}{c}{large}  \\ \cline{4-7} \cline{9-12}
&$p$ & $n$ & $z$ & CPU & Gap (\%) & \# SOLVED &  & $z$ & CPU & Gap (\%) & \# SOLVED \\
    \hline
SB & 0.02 & 1200 & 23.9  & 6.52 &  0 & 10 &  &  23.9 & 55.84 & 0 & 10  \\
&& 3600 & 72.7  & 24.42 &  0 & 10 &  & 72.7 & 258.60 &  0 & 10  \\
&& 6000 & 168.9  & 766.64 & 11.17 & 10 &  & 122.1  & 410.82 & 0 & 9  \\
&& 8400 & 234.2  & 1088.18 &  7.86  & 10 &  & 168.8 & 438.66 & 0 & 7 \\
&0.05 & 1200 &  75.2  & 330.79 & 18.42 & 10 &  & 109.0  & 6542.20 & 35.10  &  10  \\ 
&& 3600 & 269.9  & 2094.05 & 27.82 & 10 &  & 480.5  & 74868.29 & 62.15 & 5  \\
&& 6000 & 628.5  & 7330.18 & 50.52 & 10 &  &  -- & --  & --   & 0   \\
&& 8400 & 999.0 & 13140.1 & 57.29 & 10 &  & -- & --  & --   & 0  \\
AB &0.02 & 1200 & 23.9 & 6.25 & 0 & 10  &  & 23.9 & 8.77 & 0 & 10 \\
&& 3600 & 72.7 & 24.78 & 0 & 10 &  & 72.7 & 38.88 & 0 & 10 \\
&& 6000 & 121.0 & 42.41 & 0 & 10 &  & 121.0 & 67.53 & 0 & 10 \\
&& 8400 & 169.9 & 61.40 & 0 & 10 &  & 169.9 & 98.08 & 0 & 10 \\
&0.05 & 1200 &  56.3 & 10.24 & 0 & 10 &  & 56.3 & 155.94 & 0 & 10  \\ 
&& 3600 & 217.6 & 674.08 & 11.35 & 10 &  & 175.9 & 1498.80 & 0.68 & 8  \\
&& 6000 & 369.1 & 845.82 & 16.20 & 10 &  & 300.3 & 5067.35 & 0.86 & 10  \\
&& 8400 & 616.2 & 2806.38 & 25.81 & 10 &  & 432.5 & 9445.67 & 1.23 & 8  \\
\hline
\end{tabular}
\end{center}
\end{onehalfspace}
\vspace{2mm}

We also observe that, at $p = 0.05$ increasing the window size improves the gap values in contrast to CW decoder. In FW decoder, although the window size does not depend on $n$, gap values still depend on $n$ due to error accumulation during the iterations. That is, if a window is not decoded optimally, this near optimal window solution will propagate to the upcoming window decodings. As the code length $n$ gets larger, this effect becomes more apparent and the gap values increases. If the window size is larger, then we are considering more information during the window decoding, which improves the gap values. This effect is explained graphically in Figure 14.

From Table \ref{tab:SBABRW}, we can see that RW cannot complete decoding at all cases. RW decoder stores $m$--CPLEX models in memory and CPLEX needs additional memory for branch--and--bound tree while solving the window model. Hence, when the window size gets larger, we see that memory is not sufficient to complete the iterations for some instances.   

Comparison of Tables \ref{tab:SBABFW} and \ref{tab:SBABRW} shows that ABFW and ABRW methods give similar gap values as expected. However, ABRW method requires more time to manage window models. As the window size gets larger,  the computational time of ABRW is even worse than EMD (in Table 3) with high error probability.




\subsection{LDPC--C Code Results}

We also investigate the performance of FW and RW decoders for very large code length. For this purpose we take $n = 12000$ and consider high ($p = 0.05$) error probability, small and large window sizes. CW method is inapplicable in practice for very large code lengths, since it includes all the bits of the codeword as a decision variable to the window model. Performance of EMD for $n = 12000$ is given in the last row of Table \ref{tab:CPLEX--LowHighErrorSC}.

\begin{onehalfspace}
\begin{center}
\footnotesize
\captionof{table}{Performances of FW and RW decoders}
    \label{tab:CCFWRW}
\begin{tabular}{ccccccccccc}
    \hline
  & $w$ & \multicolumn{4}{c}{small} & & \multicolumn{4}{c}{large}  \\ \cline{3-6} \cline{8-11}
 & & $z$ & CPU & Gap (\%) & \# SOLVED & & $z$ & CPU & Gap (\%) & \# SOLVED \\
    \hline 
FW & SB &  926.6 & 5796.79 & 30.93 & 10 & & 597.6 & 14749.46 & 0.77 & 10\\
 & AB & 990.2 & 3686.91 & 34.03 & 10 & & 597.6 & 3999.73 & 0.77 & 10  \\
RW & SB & 1434.1  & 20013.34 & 58.21 & 10 & & -- & -- & -- & 0 \\
 & AB & 907.0 & 4537.75 & 27.74 & 10 & & 596.4  & 5428.56 & 0.45 & 5 \\
\hline
\end{tabular}
\end{center}
\end{onehalfspace}
\vspace{2mm}

Table \ref{tab:CCFWRW} summarizes the average results of 10 instances for FW and RW decoders with SB and AB alternatives. When we have small window size, all methods can decode the received vector. Among all, ABFW completed decoding within shortest time.

When the window size gets larger, RW decoder cannot solve all instances due to memory limit. On the other hand, FW decoder can solve all instances with better gap values compared with the small window size. ABFW takes less time by making use of integral starting solution advantage over SBFW. Moreover, compared with the EMD (last row of Table \ref{tab:CPLEX--LowHighErrorSC}), ABFW finds near optimal solutions in shorter time for all instances. However, EMD can solve only 2 instances to optimality. For the 5 instances that ABRW can decode, ABFW and ABRW get the same objective values. For these cases, ABFW is faster than ABRW as expected.

Considering the computational results for LDPC--C codes, we can see that ABFW is the best alternative for decoding process in terms of both time and solution quality. We further evaluate the performances of the methods by analyzing their decoding errors with respect to the original vector as given in Figure 14.

\begin{minipage}{\linewidth}	
	\centering
	\includegraphics[width=0.8\columnwidth]{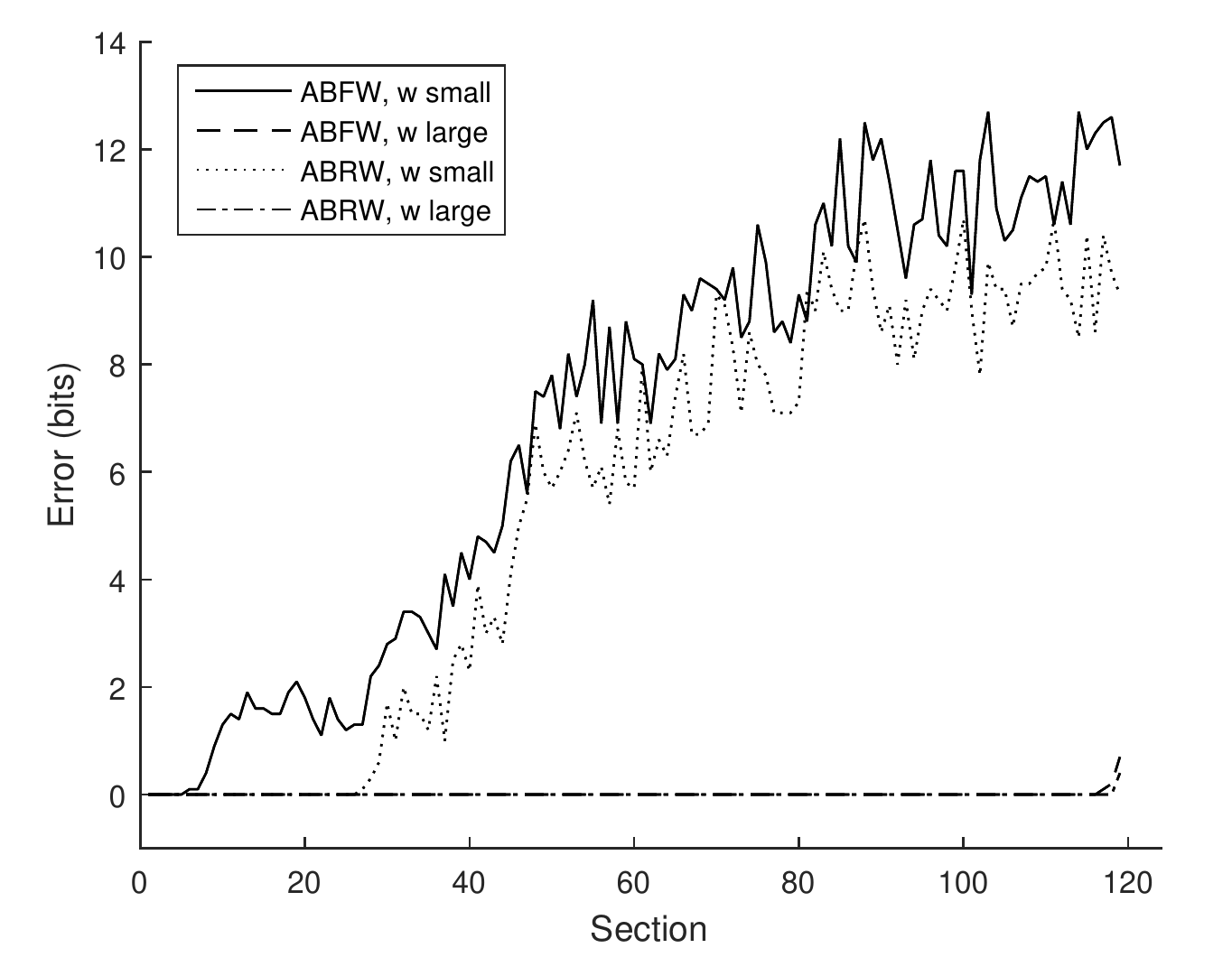}\\
	Figure 14: Error accumulation in decoding
\end{minipage}\\

In this figure, the average decoding errors of the 10 instances for code length $n = 12000$ with error probability $p = 0.05$ are given. We divide $n$ into 120 sections each include 100 bits. For each section, average errors from the original code vector is plotted.  When the window size is small, average error gets larger as the iterations proceeds. That is when we make error in decoding in early steps of the decoding process, this error will increase the probability that we are decoding erroneously in the upcoming windows. On the other hand, when the window size gets larger, we have more information about the LDPC--C code, which decreases the error accumulation during the iterations. However, taking a large window size requires more decoding time. As a result, one should take into account the trade off between computational time and the solution quality when deciding on the window size. 

The performance of decoding algorithms are interpreted with Bit Error Rate (BER) in telecommunications literature. BER is the percentage of the decoded bits that are different than the original vector \cite{M03}.

\begin{equation} \label{BER}
BER = \frac{\sum_{i=0}^{n}\mid y^{o}_{i} - y^{d}_{i}\mid}{n} \times 100
\end{equation}

BER can be calculated with the formula given in equation (\ref{BER}), where $\mathbf{y^{o}}$ is the original and $\mathbf{y^{d}}$ is the decoded codeword. 

\begin{onehalfspace}
\begin{center}
\captionof{table}{BER of Sliding Window Decoders}
    \label{tab:BERSlidingWindowDecoders}
\begin{tabular}{ccc}
    \hline
     w & ABFW &  ABRW  \\ \cline{2-3}
    \hline 
small & 6.918 & 5.402  \\ 
large & 0.008 & 0.003  \\ 
\hline
\end{tabular}
\end{center}
\end{onehalfspace}

We can calculate the BER values for our decoding algorithms using the data of Figure 15. The BER results given in Table \ref{tab:BERSlidingWindowDecoders} show that the error correction capability increases when we have larger window. For example, among 100 bits of the codeword that is decoded by ABFW method, approximately 7 bits  (\% 6.918) are different from the original codeword when window size is small. As the window size gets larger, this difference drops to 8 bits among 100,000 bits (\% 0.008).




In our final experiment, our goal is to compare our proposed decoding algorithms with two commonly used algorithms. In practical applications, decoding of a received vector is done with iterative algorithms. Among these Gallager A and B algorithms are popular due to their ease of application \cite{L05, MV10}. The performance of our proposed decoding algorithm (ABFW) can be tested against a sliding window decoder that uses Gallager A or B algorithm for decoding windows (see Algorithm 1). 


Gallager A and B algorithms are quite similar. For each bit of the received codeword $\mathbf{\hat{y}}$, the algorithm collects messages, which are the values of the parity--check equations, from each check node. If the neighboring check node is unsatisfied, then this is considered as an indication of an error in the corresponding bit. If most of the neighbors of a bit are unsatisfied, we have a strong intuition that the bit is erroneous. Let $d_i$ be the number of neighbors of  variable node $i$ in the Tanner graph of the code. 

As given in Algorithm 3, Gallager A algorithm prefers to flip the bit that has the maximum number of unsatisfied checks. At each iteration of the algorithm, we flip only one bit which guarantees that the number of unsatisfied check nodes will decrease at each iteration. Gallager B algorithm decides whether to flip or not each bit at an iteration. For each bit, Gallager B flips the bit if the number of unsatisfied check nodes is larger than the satisfied ones. In Gallager B algorithm, decrease in the unsatisfied check nodes at each iteration is not for sure since it applies multi--flip at an iteration.

\vspace{3mm}

\begin{onehalfspace}
\begin{center}
\footnotesize
$
\begin{tabular}{ll}
\textbf{Algorithm 3:} (Gallager A and B) \\
\hline
\vspace{-4mm}\\
\textbf{Input:} Received vector, $\mathbf{\hat{y}}$\\
\hline
\vspace{-4mm}\\
1.  Calculate all parity--check equations \\
2. \textbf{If} all check nodes are satisfied, \textbf{Then} STOP.\\ 
3. \textbf{Else} Calculate the number of all unsatisfied parity--check \\
\hspace{40pt} equations for all received bits, say $u_i$ for bit $i$. \\
4 - A. \hspace{7pt} Let $l = \argmax_i\{u_i\}$. \textbf{If} $u_l > d_l / 2$, \textbf{Then} flip bit $l$.\\
4 - B. \hspace{7pt}  \textbf{If} $u_i > d_i / 2$, \textbf{Then} flip bit $i$.\\ 
5. \textbf{End If} \\
6. \textbf{If} stopping is satisfied, \textbf{Then} STOP.\\ 
7. \textbf{Else} Go to Step 1.\\
8. \textbf{End If} \\
\hline
\vspace{-4mm}\\
\textbf{Output:} A feasible decoded codeword, or no solution\\
\hline
\end{tabular}
$
\end{center}
\end{onehalfspace}
\vspace{5mm}

We apply Gallager algorithm at each window of the sliding window algorithm instead of solving window model with CPLEX. A known problem with these algorithms is that they may get stuck when there is a cycle in the LDPC code \cite{SPT14}. In such a case, the algorithm may terminate with no conclusion. To avoid such a situation, we take the stopping criterion as the number of iterations and bound it with value 100. Note that this may result in ending with an infeasible solution when the algorithm terminates. 


\begin{onehalfspace}
\begin{center}
\footnotesize
\captionof{table}{Performance of Gallager A}
    \label{tab:GA}
\begin{tabular}{cccccccccccccc}
    \hline
 &  $w$  & \multicolumn{5}{c}{small} &  & \multicolumn{5}{c}{large}  \\ \cline{3-7} \cline{9-13}
$p$ & $n$ & $z$ & CPU & Gap (\%) & \# FEAS & BER &  & $z$ & CPU & Gap (\%) & \# FEAS & BER   \\
    \hline
0.02 & 1200 & 159.2 & 10.69 & 84.99 & 0 & 11.94 & & 234.1 & 20.79 & 89.79 & 0 & 18.23 \\
& 3600 & 213.1 & 49.32 & 65.99 & 0 & 4.33 &  &  276.9 & 99.38 & 73.81 & 0 & 5.89 \\
& 6000 & 268.3 & 99.14 & 54.96 & 0 & 2.81 &  &  338.2 & 192.15 & 64.27 & 0 & 3.74 \\
& 8400 & 323.9 & 159.23 & 47.63 & 0 & 2.21 &  &  386.4 & 299.31 & 56.12 & 0 & 2.72 \\
& 12000 & 395.1 & 261.76 & 39.65 & 0 & 1.67 &  &  451.5 & 506.70 & 47.22 & 0 & 1.85 \\
0.05 & 1200 & 191.3 & 11.44 & 70.62 & 0 & 15.43 &  & 259.1 & 20.74 & 78.31 & 0 &19.7 \\
& 3600 & 348.9 & 52.08 & 49.66 & 0 & 10.57 &  & 387.5 & 99.06 & 54.69 & 0 & 10.13 \\
& 6000 & 518.7 & 102.85 & 42.59 & 0 & 9.96 &  & 539.3 & 192.02 & 44.71 & 0 & 8.78 \\
& 8400 & 684.8 & 163.59 & 38.83 & 0 & 9.64 &  & 684.8 & 299.04 & 38.74 & 0 & 7.99 \\
& 12000 & 917.7 & 252.25 & 35.35 & 0 & 9.22 &  & 879.1 & 485.53 & 32.39 & 0 & 7.11 \\
\hline
\end{tabular}
\end{center}
\end{onehalfspace}
\vspace{2mm}

Table \ref{tab:GA} shows the average of 10 instances with Gallager A algorithm when it is applied in the windows of sliding window decoder. Gallager A algorithm cannot find a feasible solution for any of the cases, as given in ``\# FEAS" column.  That is the decoded vector does not satisfy the equality $\mathbf{v}\mathbf{H}^\textrm{T}=\mathbf{0}$ (mod 2). Besides, decoded vectors are far away from the best known lower bounds (found by CPLEX while obtaining the results in Table \ref{tab:CPLEX--LowHighErrorSC}) which can be seen from the ``Gap (\%)" column. 

``BER" column shows the percent difference from the original codeword. When the values compared with the ones in Table \ref{tab:BERSlidingWindowDecoders} for $n = 12000$ and $p = 0.05$, our proposed ABFW algorithm provides significantly higher quality solutions compared to Gallager A. 

\begin{onehalfspace}
\begin{center}
\footnotesize
\captionof{table}{Performance of Gallager B}
    \label{tab:GB}
\begin{tabular}{cccccccccccccc}
    \hline
 &  $w$  & \multicolumn{5}{c}{small} &  & \multicolumn{5}{c}{large}  \\ \cline{3-7} \cline{9-13}
$p$ & $n$ & $z$ & CPU & Gap (\%) & \# FEAS & BER &  & $z$ & CPU & Gap (\%) & \# FEAS & BER   \\
    \hline
0.02 & 1200 & 174.2 & 10.75 & 86.15 & 0 & 13.55 &  & 469.7 & 21.19 & 94.69 & 0 & 38.67 \\
& 3600 & 781.1 & 50.46 & 85.63 & 0 & 20.96 &  & 1645.2 & 100.27 & 95.53 & 0 & 45.51 \\
& 6000 & 1854.7 & 93.83 & 92.02 & 0 & 30.49 &  & 2846.3 & 193.40 & 95.74 & 0 & 47.34 \\
& 8400 & 3037.8 & 149.39 & 94.11 & 0 & 35.85 &  & 4056.3 & 301.41 & 95.81 & 0 & 48.19 \\
& 12000 & 4816.4 & 250.38 & 94.95 & 0 & 39.95 &  & 5865.5 & 489.15 & 95.93 & 0 & 48.83 \\
0.05 & 1200 & 519.4 & 11.63 & 89.01 & 0 & 42.87 &  & 591.4 & 21.11 & 90.48 & 0 & 49.29 \\
& 3600 & 1704.5 & 47.87 & 89.69 & 0 & 47.22 &  & 1791.3 & 100.43 & 90.19 & 0 & 49.69 \\
& 6000 & 2889.0 & 94.72 & 89.69 & 0 & 48.09 &  & 2983.8 & 194.52 & 90.02 & 0 & 49.88 \\
& 8400 & 4073.6 & 150.02 & 89.71 & 0 & 48.49 &  & 4184.2 & 302.27 & 89.99 & 0 & 50.03 \\
& 12000 & 5847.9 & 251.29 & 89.85 & 0 & 48.71 &  & 5973.1 & 489.25 & 90.07 & 0 & 49.96 \\
\hline
\end{tabular}
\end{center}
\end{onehalfspace}

As summarized Table \ref{tab:GB}, BER values are high since 
 on the contrary to Gallager A algorithm, Gallager B does not guarantee to decrease the error as its iterations proceed. That is error accumulation effect appears in BER results more dramatically for Gallager B. Both Gallager A and B algorithms are faster than ABFW method. However, their solutions are usually not feasible and are distant from the best known lower bound. 

These results indicate that ABFW is a strong candidate for decoding problem in communication systems. Gallager A and B algorithms give quick but poor quality solutions. These algorithms may be practical for TV broadcasting and video streams since fast decoding is crucial for these applications. On the other hand, as in the case of NASA's Mission Pluto, we may have some received information that cannot be reobtained from the source. For such cases high solution quality is the key issue instead of decoding speed. Hence, ABFW method is more practical for these kind of communication systems.

\section{Conclusions} \label{Conclusions}

We proposed optimization--based sliding window decoders for terminated LDPC--C codes, namely complete window (CW), finite window (FW), repeating windows (RW) decoders. We explained how one can utilize these algorithms to practically decode infinite dimensional LDPC--C codes and introduce LDPC convolutional code (CC) decoder. The computational results indicate that within the given time limit sliding window decoders find better feasible solutions in shorter time compared with exact model decoder (EMD). For each proposed decoder, we implement some binary (SB) and all binary (AB) variants. Among the sliding window decoders, AB approach is better than SB due to starting solution advantage.  

For the decoding of convolutional codes, our proposed ABFW algorithm is the best among all methods in terms of both computational time and solution quality. One can obtain better solutions by increasing the window size in the expense of computational time. 

Although, RW approach reveals worse performance than FW method, it can still be a nice candidate to decode time invariant LDPC--C codes where all windows are same. In such a case, one needs to store a single window model instead of $m$. This can decrease the memory usage and improve the computational time. 

Gallager A and B algorithms are popular in practical applications. Compared with ABFW approach, these algorithms give poor quality solution in shorter time. Our proposed algorithm ABFW can contribute to the communication system reliability by providing near optimal decoded codewords. It is applicable in settings such as deep space communications where obtaining a high--quality decoding within reasonable amount of time is crucial.

\section*{Acknowledgements} 
This research has been supported by the Turkish Scientific and Technological Research Council with grant no 113M499.


\vspace{-5mm}

\end{document}